\documentclass[submission, LectureNotes]{SciPost}

\begin{document}

\begin{center}{\Large \textbf{
An improved formulation of Jaccard's theory of the electric properties of ice
}}\end{center}

\begin{center}
J. De Poorter\textsuperscript{1}
\end{center}

\begin{center}
{\bf 1} Horatio vzw - Koningin Maria Hendrikaplein 64d, Ghent, 9000, Belgium
\\
* john.zarat@gmail.com
\end{center}

\begin{center}
\today
\end{center}


\section*{Abstract}
{\bf
In the standard derivation of Jaccard's theory of the electric properties of ice, no fundamental distinction is made between bound and free charges. This leads to some didactical problems like the ad hoc introduction of the so-called 'configuration vector' $\bf{\Omega}$. However, when the two types of charges are distinguished, it becomes clear that $\bf{\Omega}$ is redundant and proportional to the polarisation density. We also show that Jaccard's formulation contains a wrong formula for the electric susceptibility and that the correct $\Phi$ factor can be derived from a straightforward kinetic approach (which Jaccard failed to do). }

\vspace{10pt}
\noindent\rule{\textwidth}{1pt}
\tableofcontents\thispagestyle{fancy}
\noindent\rule{\textwidth}{1pt}
\vspace{10pt}

\section{Introduction}
\label{sec:intro}

Jaccard's theory provides a solid basis for the the electric properties of ice~\cite{RN13131}. The theory assumes the presence two types of free-moving defects in ice. The first type is ionic consisting of H$^+$ and OH$^-$ ions and the second type is related to wrongly oriented water molecules inside the ice lattice, the so-called Bjerrum defects~\cite{RN10943, RN10877}. The electric currents associated with the movement of these defects were described quantitatively resulting in an analytical model for the whole dielectric spectrum of ice. 

The most popular and widely cited formulation of Jaccard's theory was examined. This formulation is found in the standard work 'The physics of ice'~\cite{RN10943} and is based on the founding papers of both Jaccard~\cite{RN13131} and Hubmann~\cite{RN10877}. In order to solve the equations, the configuration vector $\Omega$ is introduced which plays a central role in the formulation. Although this vector is well defined its real physical meaning remains obscure. 

Although the polarisation density $\mathbf{P}$ plays an essential role in the whole model, no fundamental distinction is made between bound charges and free charges. However, the difference between free and bound charges is essential in solving the macroscopic Maxwell equations correctly~\cite{RN13018} and results in a simple proportionality between  $\mathbf{ \Omega}$  and the polarisation density, i.e. $\mathbf{P} = -e_{DL}\mathbf{ \Omega}$ with $e_{DL}$ the charge of the Bjerrum defects. Petrenko and Withworth~\cite{RN10943} use this proportionality in order to calculate the susceptibility but only in the case of one kind of defect (see section 4.5.1). In the general case of more than one defect (see section 4.5.2), a more complex relation is assumed. Also Jaccard~\cite{RN13131} finds only the proportionality in the special case that the Bjerrum defects dominate the conductivity. 

The main idea of this paper is that these problems arise from the confusion between bound and free charges and that the introduction of the configuration vector is unnecessary. Klyuev and others~\cite{RN11906} address this difference between bound and free charges and define the mobile defects in ice as quasi particles. However, they use a more general formulation for the dynamics of the quasi particles thereby circumventing the problems of the standard formulation. It will be shown that the currently used formula for the electric susceptibility of ice is incorrect.

\section{Reformulation of Jaccard's theory}
In this section, the theory of Jaccard is reformulated without the introduction of the configuration vector. For this purpose, the macroscopic Maxwell equations are introduced and how they define the difference between bound and free charges is discussed. 

\subsection{Defects in ice}
\begin{figure}
\centering{\includegraphics[width=86mm]{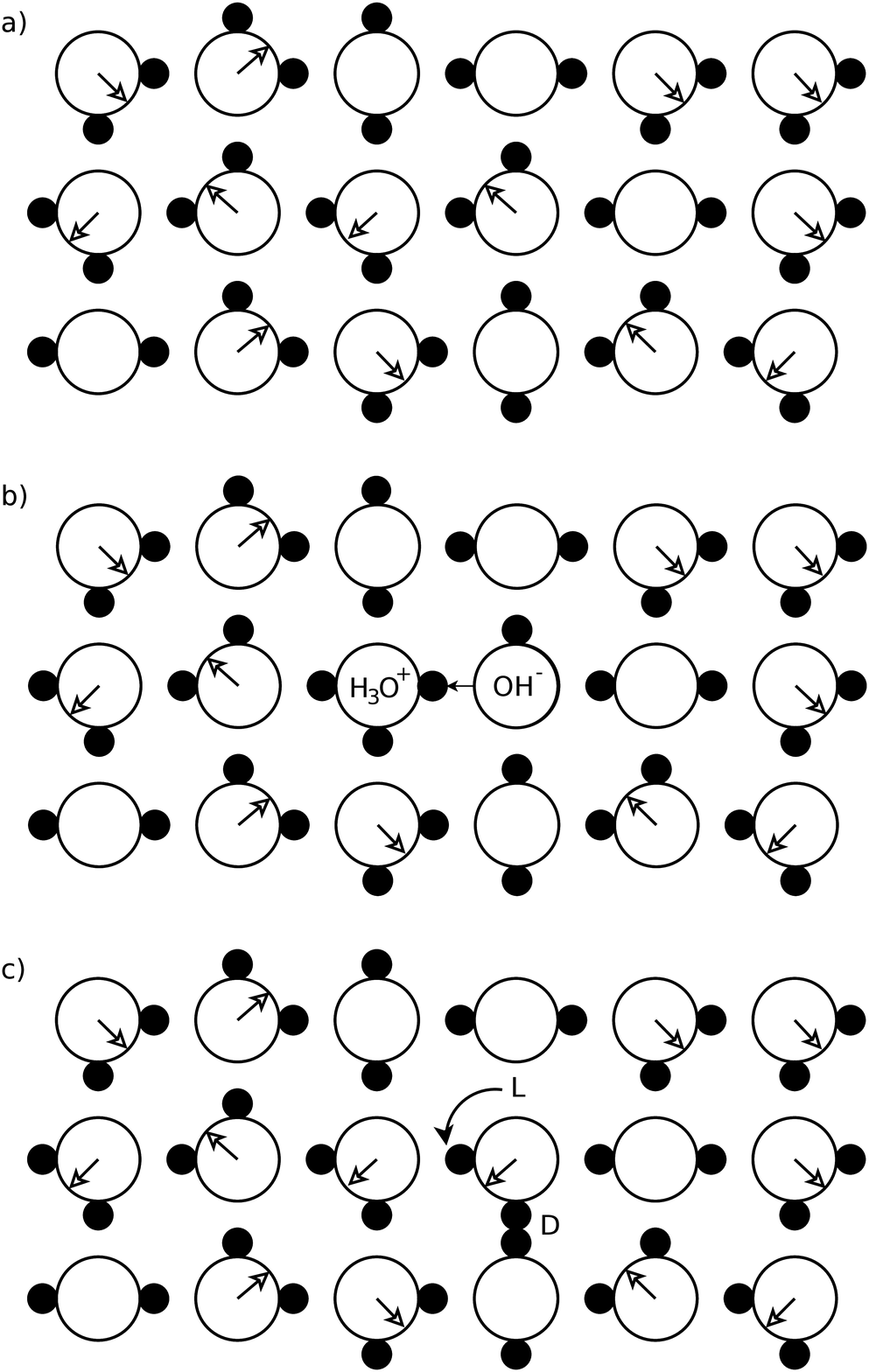}}
\caption{ 2D representation of the ice lattice. The white circles are oxygen atoms, the black ones hydrogen atoms. a) A defect-free ice lattice, b) an H$^+$ and OH$^-$ defect pair appears when a hydrogen nucleus jumps to a neighbouring water molecule and (c) a Bjerrum D and L defect pair appears when one of hydrogen atoms is positioned differently.  The dipole moment of the individual water molecules $\mathbf{p_o}$ is indicated by the small arrows inside the oxygen atoms.}
\label{fig:2Dmodel1}
\end{figure}

The most common crystalline phase of ice, ice I$_h$, is hexagonal. Each water molecule is fixed in the ice crystal structure and surrounded by 4 nearest neighbours located at the corners of a tetrahedron~\cite{RN10913}. There are two rules describing the orientation of individual water molecules in the crystal structure, the so-called Bernal-Fowler ice rules. The first rule is that each molecule accepts two hydrogen atoms from two nearest-neighbours water molecules and also donates two hydrogen atoms to the two other nearest-neighbours. The second rule states that there is precisely one hydrogen atom between each nearest-neighbours pair of oxygen atoms. As an illustration of these rules, we have drawn a 2D version of a defect-free ice lattice in Fig.~\ref{fig:2Dmodel1}(a). The chosen 2D lattice differs from the ones generally chosen in literature where the angle between the hydrogen atoms is fixed at 90$^\circ$ \cite{RN10943, RN10914}. These fixed 2D representations have only 4 possible orientations of the water molecules, while in the 3D ice lattice, there are 6. Therefore, we chose a 2D representation which contains also water molecules with hydrogen atoms positioned opposite to each other. The water molecules in Fig.~\ref{fig:2Dmodel1}(a) have also 6 possible orientations. This allows more randomness in the 2D lattices corresponding better to the entropy of the real 3D lattices. Notice that an ice crystal has a lack of long-range order in the orientation of the water molecules, a property essential to understand its electric properties. In the absence of an external electric field the net polarisation induced by the permanent dipoles is zero due to this lack of long-range order. In our 2D representation one third of the molecules, the ones with the hydrogen atoms opposite to each other, have no netto dipole moment. In 3D, all the water molecules have a dipole moment. But if the plane of interest is well chosen the dipole moment of one third of the molecules is also oriented perpendicular to this plane, leading to configurations similar as in the 2D lattice~\cite{RN10877}. 

When an electric field $E$ is applied to ice, the ice becomes easily polarised and has a large relative dielectric constant (93 at {-}$3^\circ$C)\cite{RN8152}. Only a small fraction of this constant can be explained by the polarisation of the charge distribution of individual water molecules, the majority of this effect is due to the reorientation of the individual watermolecule dipoles in the direction of the applied field. However, this reorientation can only happen if the ice rules are at least temporary broken. Therefore, mobile lattice defects are assumed to be present inside the ice crystal changing the orientations of the water molecules when they move. 

Two types of defects are described in ice: H$^+$-OH$^-$ ionic defect pairs which violate the first ice rule and the Bjerrum defect pairs embodying the violation of the second ice rule~\cite{RN10943}. A H$^+$-OH$^-$ ionic defect pair is created when one of the hydrogen atoms jumps to the neighbouring water molecule leaving its electron behind (see Fig.~\ref{fig:2Dmodel1}(b)). Both ions can separate and move independently throughout the lattice. The second ice rule is violated by turning one watermolecule over an angle of 90$^\circ$ so that one of its neighbouring O-O bonds is occupied by two hydrogen atoms (a Bjerrum D defect) and the other one with no hydrogen atoms (a Bjerrum L defect) (see Fig.~\ref{fig:2Dmodel1}(c)). The Bjerrum defects can also separate and move independently throughout the lattice. They are seen as quasi particles because they behave in a way similar to a real charged particle \cite{RN11906}. However, they are not real physical particles, but only a temporary deviations of the Bernal-Fowler rule in the ice structure. 

\begin{figure*}
\centering{\includegraphics[width=110mm]{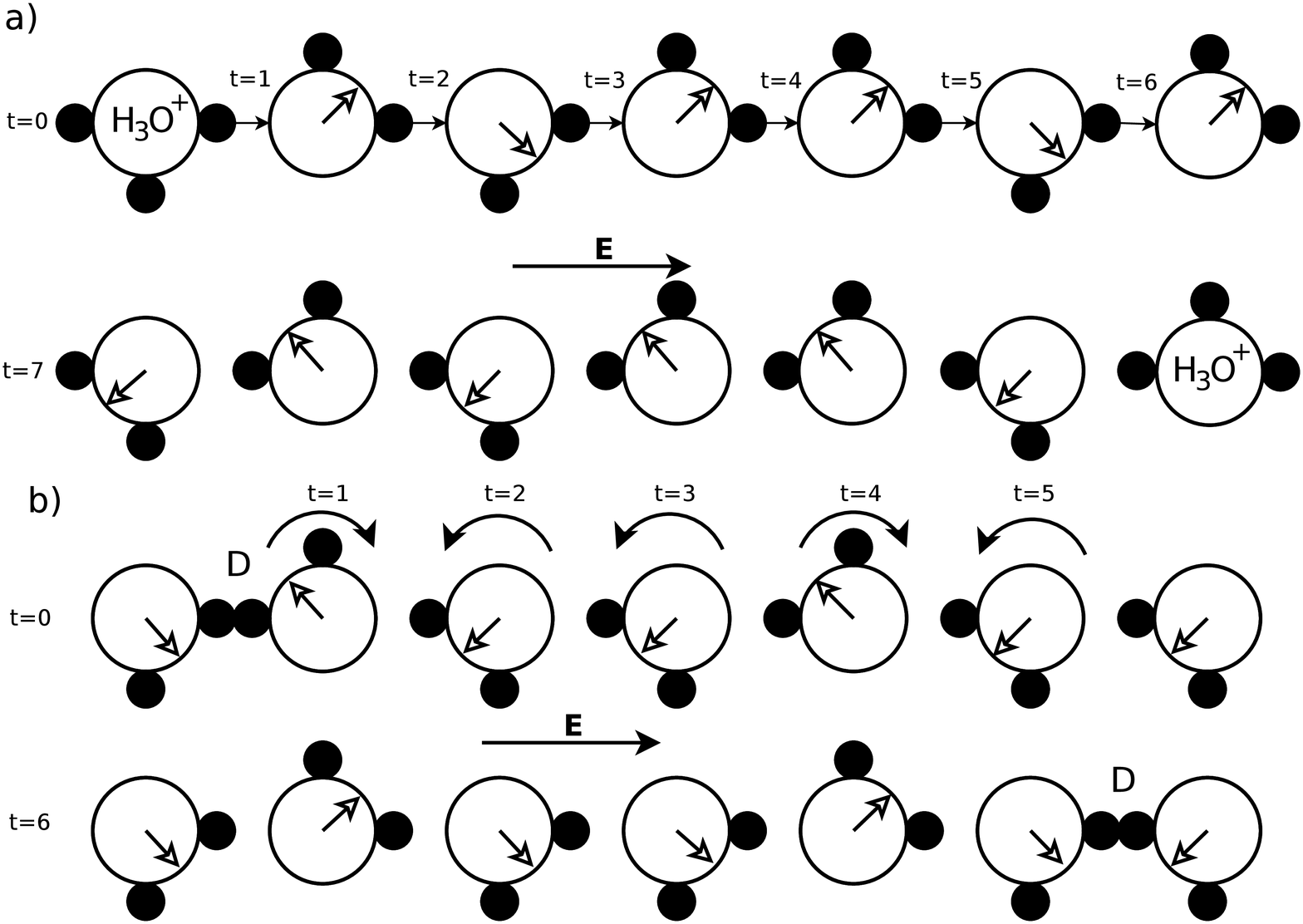}}
\caption{2D representation of the movement of defects in ice under the influence of an external electric field $\bf{E}$. The dipole moment of the individual water molecules $\mathbf{p_o}$ is indicated by the small arrows inside the oxygen atoms. First row: a H$^+$ defect (a) or a D defect (b)  is present at the left of a well-chosen chain of water molecules at t = 0. Second row: the orientation of the water molecules after the movement.} 
\label{fig:2Dmodel2} 
\end{figure*}    

In Fig.~\ref{fig:2Dmodel2} we visualised the movement of the H$^+$ ion and the D defect under the influence of an external electric field. In the upper row of (a), the H$^+$ ion jumps through a well-chosen chain of water molecules from one molecule to the next, the so-called Grotthus mechanism~\cite{RN11914}. Notice that these consecutive jumps result in the lower chain consisting of  water molecules with changed dipole moments. For a H$^+$ ion to pass through a chain, the water molecules in front of the ion must have a dipole moment with a component in the same direction as the electric field.  As a consequence, the passing of one H$^+$ prevents other H$^+$ ions to move in the same path. 

In Fig.~\ref{fig:2Dmodel2}b (upper row), the water molecules allowing a D defect to pass by are  oriented in the direction opposite to the electric field. The driving force for this movement is the netto torque on the water molecules trying to orient themselves with a dipole moment parallel to the electric field. So the passing by of a D defect from left to right, will also change the orientation of the water molecules and also here, the D defect can travel this path only once. However, the new chain is now open for moving H$^+$ (and OH$^-$) ions, just like the movement of a H$^+$ ion opens the chain for D (and L) defect transport. This property of the defects is essential for understanding how defects interact with each other. In a similar way one can show that an L defect opens the path for both ions and vice versa the OH$^-$ ion opens the chain for both Bjerrum defects.   

\begin{figure*}
\centering{\includegraphics[width=115mm]{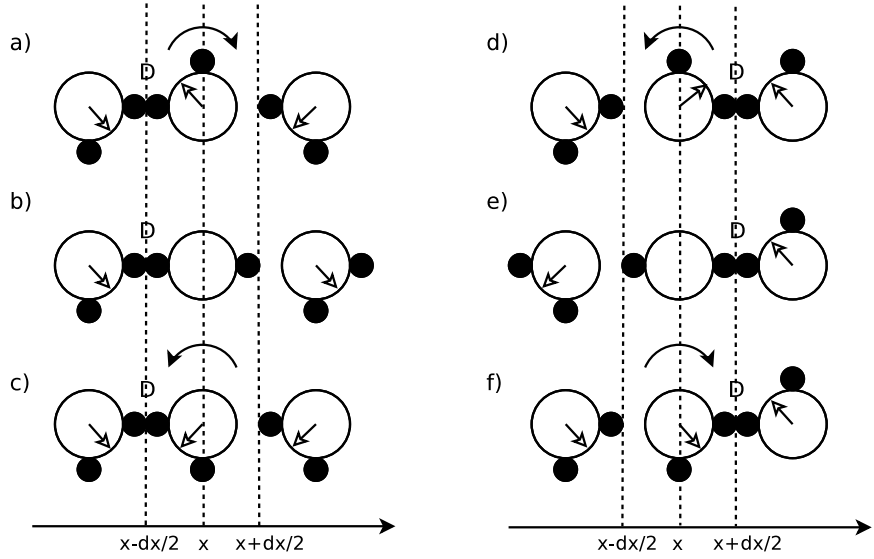}}
\caption{D defects may move horizontally in an 2D ice-like lattice. In a-c the D defect can jump to the right side following the turning indicated by the arrows, however in b the D defect is blocked in this direction. In d-f the three possible jumps to the left are shown, only e is blocked. } 
\label{fig:current}
\end{figure*}  

In Fig.~\ref{fig:2Dmodel2}b the chains are ideally ordered to obtain a long range traveling through the lattice. However, This is not a realistic ordening, a normal ice lattice is chaotically ordered (like in Fig.~\ref{fig:2Dmodel1}). In Fig.~\ref{fig:current}a-c, the three possible orientations a D defect may encounter moving to the right in a 2D lattice are drawn. Those three possible orientations, obey the ice rules, and are as likely to occur. It is clear that in b the movement of the D defect to the right is blocked. So, in one third of its jumps to the right a D defect will be blocked in its movement. The orientations drawn in d-f are for a D defect travelling to the left. A similar blocking is seen in e. A similar figure can be drawn for the ions. Their movement in a certain direction will also be blocked in one third of the possible orientations.   

\subsection{Macroscopic Maxwell equations}
To describe the electric fields inside ice, the macroscopic Maxwell equations are used~\cite{RN13018}. The fields in these equations are a macroscopic spatial average of the microscopic fields around the molecules. In order to calculate these fields one has to differentiate  between two different types of charges: free charges  and bound charges. Free charges, like ions, will move freely through an ice specimen under influence of an applied electric field, while bound charges  will displace only locally. Bound charges remain connected to the individual molecules. 

In the macroscopic Maxwell formulation, free charges with a charge density $\rho_f$ are the source of the electric displacement field $\mathbf{D}$ and are related to them by the integral formulation of Gauss's law~\cite{RN13018}
 
\begin{equation}
 \iint_S \mathbf{D}.\mathbf{dS} = \iiint_V \rho_f dV \label{eq:Maxwell1} 
\end{equation} 
with $V$ the volume containing the free charge and $S$ the surrounding surface through which the netto flux of $\mathbf{D}$ is calculated. 
$\mathbf{D}$ is defined as
 
\begin{equation}
\mathbf{D} = \epsilon_o \mathbf{E} + \mathbf{P}, \label{eq:Maxwell2}
\end{equation} 
with $\epsilon_o$ the vacuum permittivity, $\mathbf{ P}$ the polarisation density and $\mathbf{ E}$ the macroscopic electric field. $\mathbf{ P}$  plays a central role in Jaccard's theory of ice and is a direct consequence of the presence of bound charges. The density of the bound charges $\rho_b$ in a volume $V$ can be calculated out the flux of the polarisation density through the surface $S$ around $V$ using~\cite{RN13018}
 
\begin{equation}
\iiint_V \rho_b dV= -\iint_S \mathbf{P}.\mathbf{dS}. \label{eq:Maxwell3}
\end{equation} 
The polarisation density $\mathbf{P}$ itself is build up by the current density of the same bound charges $\mathbf{J_b}$~\cite{RN13018} 
 
\begin{equation}
\frac{\partial \mathbf{P_l}}{\partial t} = \mathbf{J_b}.   \label{eq:PlJ}
\end{equation} 
It is this equation that is misinterpreted in Jaccard's theory~\cite{RN10943, RN10877, RN10878}. Jaccard uses Eq.~\ref{eq:PlJ} with the total current density, which is the sum of the current of the bound and free charges. The impact of this will be discussed in the next section. 
 
\begin{figure}
\centering{\includegraphics[width=86mm]{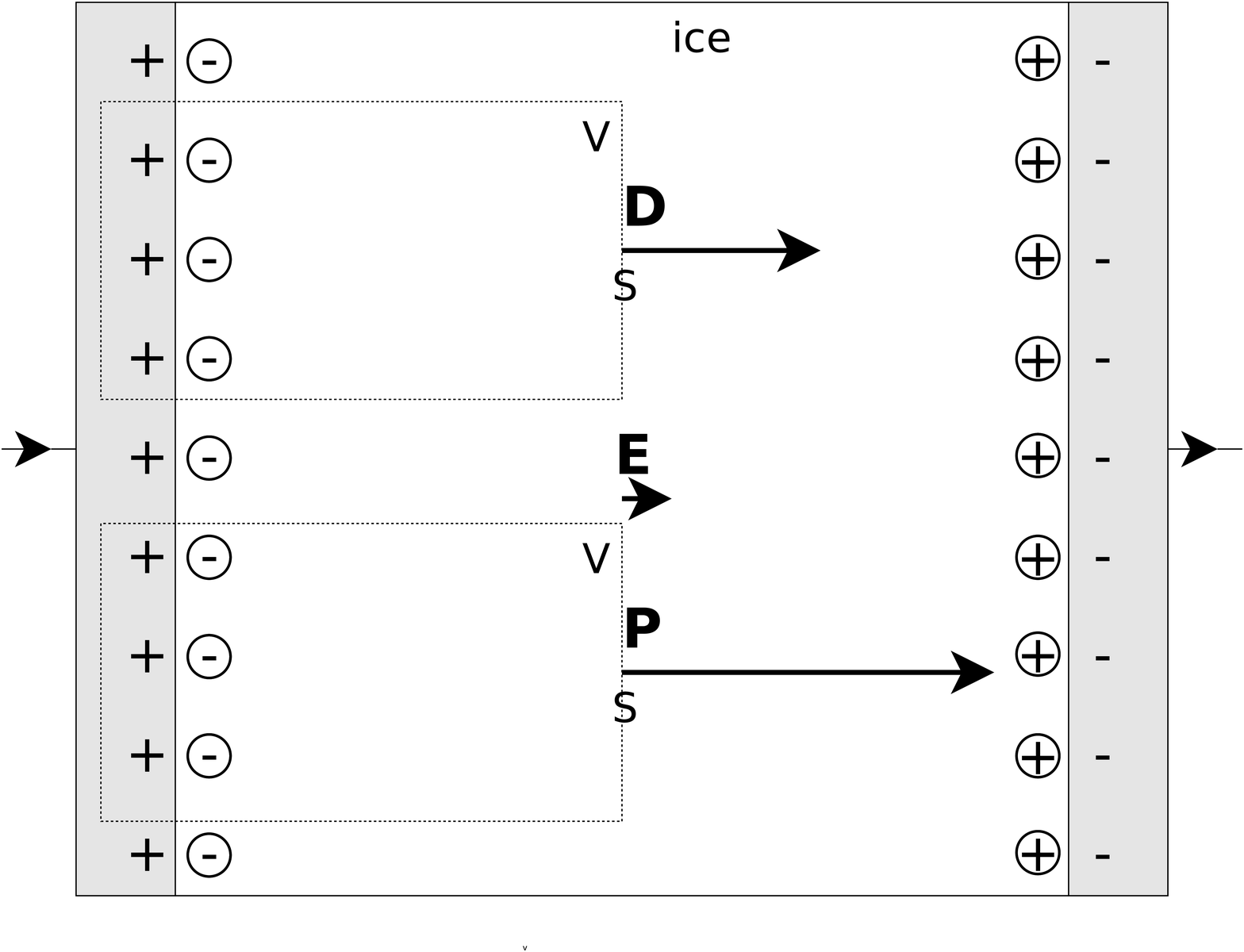}}
\caption{ A schematic view of a capacitor containing an ice slab with conductivity zero after equilibrium is reached. Both free charges at the surface of the metal contact and the bound charges at the surface of the ice slab are visualised. Also, two volumes necessary to calculate the electric displacement field $\mathbf{D}$  and the polarisation density $\mathbf{P}$ are drawn.  }
\label{fig:capacitor}
\end{figure}    

To get some insight in the macroscopic Maxwell equations we will use them in a very common case where ice is homogeneously polarised by a homogeneous electric field, induced by the plates of a parallel-plate capacitor (Fig.~\ref{fig:capacitor}). Using an external voltage source, a surface density $\sigma_f$ of free charges is accumulated at the capacitor plates. If we ignore the small amount of free charges in ice, we can calculate the electric displacement field $\mathbf{D}$ induced by the capacitor inside the ice slab out of Eq.~\ref{eq:Maxwell1}. Therefore we define a volume drawn in Fig.~\ref{fig:capacitor} that takes into account the symmetry of the setup and the fact that both $\mathbf{E}$ and $\mathbf{D}$ are zero inside the capacitor plates. We get
 
\begin{equation}
D = \epsilon_o E + P = \sigma_f.  \label{eq:cap1} 
\end{equation}  
Because the ice polarisation density $\mathbf{P}$ is homogeneous, there is no net polarisation flux through volumes defined completely in the ice, indeed  Eq.~\ref{eq:Maxwell3} leads to $\rho_b = 0$. Of course, there are microscopically bound charges, but they are averaged out macroscopically. Only at the surface, there is a shift in polarisation because the polarisation is zero inside the plates. The surface density of bound charges $\sigma_b$ can be derived applying Eq.~\ref{eq:Maxwell3} on the second volume of Fig.~\ref{fig:capacitor}, 

\begin{equation}
 \sigma_b = -P.  \label{eq:cap2} 
\end{equation}    
These bound charges decrease the electric field inside the ice slab

\begin{equation}
E = \frac{\sigma_f + \sigma_b}{ \epsilon_o }.  \label{eq:cap3} 
\end{equation}
Notice that $\sigma_b$ is negative. This equation clearly shows that the source of the electric field are both the free and the bound charges. The closer $\sigma_b$ comes to the amount of free charges $\sigma_f$, the more the free charges at the capacitors plates are compensated and the smaller the electric field is. To have an idea of the size of $\sigma_b$, we first calculate the polarisation density, which in ice is proportional to the electric field  

\begin{equation}
\mathbf{P} =  \epsilon_o \chi^{i}\mathbf{E}  \label{eq:cap4} 
\end{equation}
which $\chi^{i}$ the electric susceptibility of ice. Combing this equation with Eq.~\ref{eq:cap1} results in the more common relation that 

\begin{equation}
E = \frac{\sigma_f }{ \epsilon_o (1+\chi^{i})} = \frac{\sigma_f }{ \epsilon_o \epsilon_r^i}.  \label{eq:cap5} 
\end{equation}
where $\epsilon_r^i$ is the relative dielectric constant of ice. Together with Eq.~\ref{eq:cap3}, this leads to 

\begin{equation}
 \sigma_b = -\frac{\sigma_f \chi^{i} }{ 1+\chi^{i}}.  \label{eq:cap6} 
\end{equation}
Because ice has a large value of the dielectric constant (93 at {-}$3^\circ$C)\cite{RN8152}, the bound charges will compensate for almost 99\% of the free charges, leading to a significantly reduced electric field inside the ice slab. 

 \subsection{Bound and free charges in ice}

Real ice contains H$^+$-OH$^-$ ionic defect pairs which are free to move through the specimen. The charges of these ions are $\pm e$, but we will prove further in this text that these ions are always accompanied by oppositely charged bound charges reducing the charge that is traveled through the ice to $\pm e_{\pm}$.  

Bound charges are related to changes in $\mathbf{ P}$, the polarisation density.  It is important to distinguish between two kinds of contributions to the polarisation density. The first contribution is related to the displacement of the charge distribution of individual water molecules under influence of an external field and is called the water molecule polarisation $\mathbf{ P_m}$. $\mathbf{ P_m}$ is only a small fraction (only some procent) of $\mathbf{ P}$, the total polarisation density of ice~\cite{RN10943}. The lattice polarisation $\mathbf{ P_l}$ provides the major part of the polarisation. It is caused by the netto orientation of the permanent dipole moments $p_o$ of the water molecules. 

For simplicity reasons, we limit our reasoning to a 2D lattice. All the equations we will derive will also be valid in for an isotropic ice lattice as described by Hubman~\cite{RN10877}. We choose the horizontal x axis as the potential direction for an external field. We can project all the dipole moments of the water molecules on this axis. These components have a size called $p_{o,\parallel}$ and they can be positive or negative depending on the direction of the water molecules related to the x axis. We define $n_{p^+}$ and $n_{p^-}$  as the density of water molecules polarised in the positive and negative x direction, respectively and $n_{p=o}$ as density of the water molecules with a zero polarisation component in the x direction.  Notice that $n_o = n_{p^+} + n_{p^-}+n_{p=o}$ and that in the absence of an electric field $n_{p^+}=n_{p^-}=n_{p=o}=n_o/3$, so resulting in a zero polarisation. If an external field is applied in the direction of the positive x axis, the polarisation density of the (2D) ice lattice $\mathbf{P_l}$ in the horizontal axis is by definition equal to

\begin{equation}
\mathbf{P_l}=  (n_{p^+}-n_{p-})p_{o,\parallel}  \mathbf{e_x},   \label{eq:Pl} 
\end{equation}  
with $\mathbf{e_x}$ the unit vector in the x direction. 

\begin{figure}
\centering{\includegraphics[width=86mm]{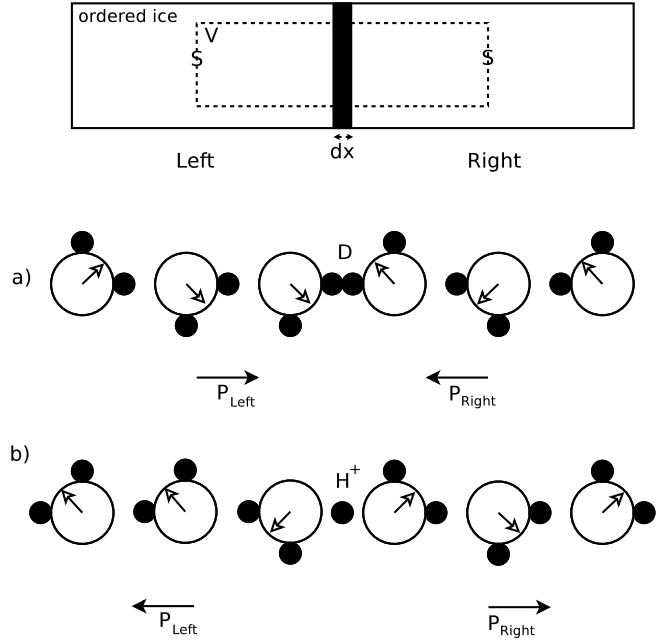}}
\caption{ An ideal specimen of ordered ice containing a one-molecule thick layer of defects and a left and right part of the specimen with water molecules oriented in accordance to the defects in that layer. The orientation of the water molecules around (a) a D defect and b) a H$^+$ ion are drawn.} 
\label{fig:boundcharges}
\end{figure}

Because both the ions and the DL defects are changing locally the orientation and netto polarisation of the water molecules, we will first quantify the bound charges that are related to these defects. To calculate the bound charge we construct an ideal ice specimen as illustrated in Fig.~\ref{fig:boundcharges} containing a one-molecule thick layer of defects (size $dx$). The left and right part of the specimen contains perfectly ordered water molecules oriented in accordance to the defects in the layer. Because the ideal uniformity of the orientation of the water molecules, both the left and the right part of the specimen have a large macroscopic polarisation density, named $ \mathbf{P_{Left}}$ and  $\mathbf{P_{Right}}$. We now define a volume $V$ and the corresponding vertical surface $S$ as in Fig.~\ref{fig:boundcharges}. This volume contains the bound charges of the layer and so we are able to solve Eq.~\ref{eq:Maxwell3} for different types of defects. 

We first consider a layer of D defects (Fig.~\ref{fig:boundcharges}a) and define $e_D^b$ as the charge of one D defect. Because there are only D defects in the central layer of thickness $dx$, we can easily fill in Eq.~\ref{eq:Maxwell3} for volume V,

\begin{equation}
e_{D}^b n_o S dx =  -(-P_{Left}-P_{Right}) S=  -(-2 p_{o,\parallel} n_o) S  , \label{eq:eDL1}
\end{equation}    
with $p_{o,\parallel}$ the component of the dipole moment of the water molecules parallel to the horizontal axis and $n_o$ the particle density of ice. This equation can be simplified as

\begin{equation}
e_{D}^b  =   \frac{2p_{o,\parallel}}{dx} = e_{DL}. \label{eq:eDL2}
\end{equation}   
A similar equation can be found for L defects, 

\begin{equation}
e_{L}^b  =  -\frac{2 p_{o,\parallel}}{dx} = - e_{DL}. \label{eq:eDL3}
\end{equation} 
proving that both charges have an opposite sign. Out of experimental data, it is found that $e_{DL}$ ($= 0.38 e$) for the D defect and $-e_{DL}$ ($= -0.38 e$) for the L defect~\cite{RN10943}. 

A similar approach also works for the ions. We now will calculate $e_+^b$ corresponding with a H$^+$ ion visualised in Fig.~\ref{fig:boundcharges}b. The total bound charge of the layer with volume $dx S$ is 

\begin{equation}
e_{+}^b n_o S dx=  -(2 p_{o,\parallel} n_o) S, \label{eq:e+1}
\end{equation}     
leading to the conclusion that  

\begin{equation}
e_{+}^b =  -\frac{2 p_{o,\parallel}}{dx} = - e_{DL}. \label{eq:e+2}
\end{equation} 
The bound charge of the H$^+$ is negative and has the same size of the D defect. The total current in ice is the sum of the currents of the free and the bound charges. This explains why a H$^+$ ion transported through the lattice carries only a charge 

\begin{equation}
e_{+} = e  - e_{DL}. \label{eq:e+3}
\end{equation} 
The displacement of the free charge $e$ is counterbalanced by that of the bound charge related to the H$^+$ ion. Another way to visualise $e = e_{+}  + e_{DL}$ is to bring an external proton (of charge $e$) inside a perfect ice lattice (like in Fig.~\ref{fig:2Dmodel1}(a)). This proton will stick on one of the oxygen atoms and will create both a H$^+$ defect (of charge $e_\pm$) and a D defect (of charge $e_{DL}$). Both will move through the lattice independently. 

Similar results can be obtained for OH$^-$,

\begin{equation}
e_{-}^b =  \frac{2 p_{o,\parallel}}{dx} =  e_{DL}. \label{eq:e+4}
\end{equation} 
Leading to the general result that 

\begin{equation}
e_{\pm} = e  - e_{DL}, \label{eq:e+5}
\end{equation} 
with $e_\pm$ the absolute value of the amount of charge that is transported by the moving ions. This equation also allows us to calculate $e_\pm$ as $0.62 e$~\cite{RN10943}. 

One of the reasons for this complete discussion is that in the Jaccard theory $e_{\pm}$ is seen as the bound charge of the ions~\cite{RN10877, RN10878}. The discussion above proves that this is not correct. The electrical charges corresponding to the different defects in ice are summarised in Table~\ref{Table1}. 

\begin{table}
\caption{ A summary of the defects in liquid water, their charges, and the current densities of both bound and free charges. }
\label{Table1}
\begin{tabular}{   c  || c   c    c   c   c}
\hline 
defect& free &bound &total  &$\mathbf{J_b}$ &$\mathbf{J_f}$\\
\hline 
H$^+$ & e& $-e_{DL}$&$e_{\pm}$ & $-e_{DL} \mathbf{j_{H^+}}$ &$e \mathbf{j_{H^+}}$\\
OH$^-$ & $-e$& $e_{DL}$& $-e_{\pm}$ & $e_{DL} \mathbf{j_{OH^-}}$ &$-e \mathbf{j_{OH^-}}$\\
D &  0& $e_{DL}$& $e_{DL}$ &  $e_{DL} \mathbf{j_{D}}$ &0\\
L &  0& $-e_{DL}$&$-e_{DL}$ &  $-e_{DL}\mathbf{ j_{L}}$ &0 \\
\hline 
\end{tabular}
\par
\end{table}

\subsection{The currents of bound and free charges}

DL defects move throughout the ice lattice. They do not contain a net physical charge, but they contain a microscopic jump in the polarisation around the defect, creating a local bound charge at the position of the defect. Because there are as many D as L defects and they are thermally created in a random way all over the ice lattice, there is no macroscopic polarisation density related to their presence. They are quasi particles because an external electric field will drive them, just like real charged particles, in an ordered way throughout the lattice (see Fig.~\ref{fig:2Dmodel2}). D defects move in the same direction as the electric field, the L defects in the opposite direction. During this movement the local polarisation is changed, creating a movement of bound charges that can be described in a similar way as the movement of a real charge. So, the DL defects are seen as a charged quasi particles with a charge, a mobility and a conductivity. The conductivity of the Bjerrum defects $\sigma_D$ and $\sigma_L$ is equal to     

\begin{equation}
\sigma_D = e_{DL} n_D \mu_D   \label{eq:s3}
\end{equation} 
and

\begin{equation}
\sigma_L = e_{DL} n_L \mu_L, \label{eq:s4}
\end{equation} 
with $\mu_D$, $\mu_L$  the positive mobilities and $n_D$, $n_L$ the particle densities of the D and L defects, respectively~\cite{RN10943}. It is important to focus on the interpretation of the different physical quantities here. $n_D$ and $n_L$ are the equilibrium densities of the defects. Because the conductivity in ice is very low, we may assume that $n_D$ and $n_L$ are significantly smaller than $n_o$~\cite{RN10943}. 
 
In a similar way, we can see the ions as quasi particles. They are a combination of a real particle (the ion) and a corresponding bound charge of the opposite sign. $\sigma_+$ is the conductivity of the positive H$^+$ ions in an unpolarised specimen,  

\begin{equation}
\sigma_+ = e_\pm n_+ \mu_+ . \label{eq:s1}
\end{equation} 
with $n_+$ the particle density and $\mu_+$ the H$^+$ mobility. For the OH$^-$ ions with conductivity $\sigma_-$, the relation is 

\begin{equation}
\sigma_- = e_\pm n_- \mu_- . \label{eq:s2}
\end{equation}
 
The flux density of the defects (i.e. the number of defects crossing a unit area per unit of time) are denoted by the vectors $\bf{j_{+}}$, $\bf{j_{-}}$, $\bf{j_D}$, $\bf{j_L}$ for H$^+$, OH$^-$ ions, D defects and L defects, respectively. These flux densities differ from the current densities (denoted with a capital letter $\bf{J_{+}}$, $\bf{J_{-}}$, ... ) and are quantifying the number of defects passing by while the corresponding current densities are quantifying the net charge. 

The flux densities are caused by two effects~\cite{RN10943}. First, an applied electric field will move the charged defects and ions through the lattice, which is described by Ohm's law. However, this movement is counterbalanced by a diffusive current of the same defects. If the lattice has no netto orientation ($P_l = 0$), there will be no netto thermal displacement of the defects. However, the ohmic currents create a dominant orientation in the water molecules ($P_l  \neq 0$) inducing a preferable direction in the thermal hopping of the defects. 

We will do the full calculation for the D-defect flux density. An electric field is applied in the horizontal direction (in the direction of $\mathbf{e_x}$). The flux density related to the electric field is easily obtained from the electric current of D defects divided by the charge transported by the D defect (see first term in Eq.~\ref{eq:fluxD1}). The second term is more complex and different ways of deriving this term are proposed in literature~\cite{RN10943, RN10877, RN10878}. We will show that this term is best seen as a simple diffusion term, related to the induced concentration gradient in $n_D$. If $D_D$ is the diffusion constant of the D defects, we get at position $x$   

\begin{equation}
\mathbf{j_D} = \frac{\sigma_D  \mathbf{E}  }{e_{DL}}  - D_D \frac{n_D(x+dx/2) -n_D(x-dx/2)}{dx}    \mathbf{e_x}. \label{eq:fluxD1} 
\end{equation}
We have drawn the different positions ($x-dx/2$, $x+dx/2$) in Fig.~\ref{fig:current}. $n_D(x+dx/2)$ is the concentration of D defects at the right side of the water molecules at x and can be calculated just using statistical considerations. The density of D defects at position $x$ is defined as

\begin{equation}
n_D(x) = \frac{n_D(x-dx/2)+n_D(x+dx/2)}{2}.
\end{equation}
Using this definition, the density of water molecules at position x involved in a D defect is $2n_D(x)$. These D defects can be located at $x-dx/2$ or at $x+dx/2$. In Fig.~\ref{fig:current}d and \ref{fig:current}f it is clearly seen that water molecules at position x are part of a D defect at the right side if they are oriented to the right and have a density equal to $n_{p^+}(x)$. Fig.~\ref{fig:current}e shows that they can also be part of a D defect at $x+dx/2$ if they have no netto dipole moment and the hydrogen atoms are oriented horizontally. These water molecules have a density equal to $n_{p=o}(x)/2$ because only half of the water molecules with a zero dipole moment is oriented horizontally. So $(n_{p^+}(x)+n_{p=o}(x)/2)/n_o$ is the fraction of the water molecules at position x that is well oriented to be part of a D defect at position $x+dx/2$. This results in following expression for $n_D(x+dx/2)$,

\begin{equation}
n_D(x+dx/2) = 2n_D(x)\frac{n_{p^+}(x)+\frac{n_{p=o}(x)}{2}}{n_o}. \label{n_Ddx}
\end{equation}
Similarly we find

\begin{equation}
n_D(x-dx/2) = 2n_D(x)\frac{n_{p^-}(x)+\frac{n_{p=o}(x)}{2}}{n_o},\label{n_Ddx2}
\end{equation}
because there are only D defects at the left side if the water molecule at x is oriented to the left side (see Fig.~\ref{fig:current}a and \ref{fig:current}c) or the hydrogen atoms of the water molecule at x are oriented horizontally (see Fig.~\ref{fig:current}b). Combining the expressions for $n_D(x+dx/2)$ and $n_D(x-dx/2)$ with Eq.~\ref{eq:Pl}, we can rewrite the flux density of the D defects as

\begin{equation}
\mathbf{j_D} = \frac{\sigma_D  \mathbf{E}  }{e_{DL}}  - \frac{2 n_D D_D}{dx n_o p_{o,\parallel} }  \mathbf{P_l}. \label{eq:fluxD2b} 
\end{equation}

This equation can also be derived for a 3D hexagonal ice I$_h$ lattice, if the small anisotropy of ice is ignored (around 15\% in the z direction~\cite{RN13136}). The hexagonal lattice is then isotropic with a tetrahedral symmetry~\cite{RN10877, RN10878}. Before we proceed, we will first summarise the properties of the ice lattice that are related to this tetrahedral symmetry. A fundamental quantity of the lattice is $r_{oo}$, the distance between two oxygen atoms of neighbouring hydrogen bonded water molecules. In tetrahedral lattice, the density of ice $n_o$ relates to $r_{oo}$ as~\cite{RN10878}

\begin{equation}
n_o = \frac{ 3 \sqrt{3} }{8r_{oo}^3}. \label{eq:no}
\end{equation}
The mean distance $dx$ between the successive planes of water molecules is equal to

\begin{equation}
dx = \frac{ 2 r_{oo}}{\sqrt{3}}, \label{eq:tetra2}
\end{equation}
This equation is obtained from $n_o = 1/dx^3$ assuming an isotropic specimen. Combining it with Eq.~\ref{eq:eDL2}, we obtain

\begin{equation}
p_{o,\parallel} = \frac{ e_{DL} r_{oo}}{\sqrt{3}}.  \label{eq:tetra3}
\end{equation}
Combining the equation for the flux density in Eq.~\ref{eq:fluxD2b} with Eqs.~\ref{eq:no}, \ref{eq:tetra2} and \ref{eq:tetra3} results in 

\begin{equation}
\mathbf{j_D} = \frac{\sigma_D  \mathbf{E}  }{e_{DL}}  -  \frac{8r_{oo} n_D D_D}{\sqrt{3}e_{DL}}  \mathbf{P_l}. \label{eq:fluxD3} 
\end{equation}  
The Einstein relation relates the diffusion coefficient of the D defects D$_D$ to the mobility of the D defects, 

\begin{equation}
D_D = \frac{k T\mu_{D}}{e_{DL}} .  \label{eq:DD}
\end{equation}
with $k$ the Boltzmann constant and T is the absolute temperature. If this relation is put into Eq.~\ref{eq:fluxD3} and using the definition of $\sigma_D$ from Eq.~\ref{eq:s3}, we obtain the well-known equation for $\mathbf{j_D}$, i.e.

\begin{equation}
\mathbf{j_D} = \frac{\sigma_D  \mathbf{E}  }{e_{DL}}  -  \Phi \frac{\sigma_D}{e_{DL}^2} \frac{\mathbf{P_l}}{e_{DL}} , \label{eq:fluxD4} 
\end{equation}  
with the $\Phi$ factor equal to

\begin{equation}
\Phi =   \frac{8 r_{oo} kT}{\sqrt{3}}. \label{eq:fluxD5} 
\end{equation}  

This $\Phi$ factor corresponds with the experimental values and is found to be independent of the anisotropy of the ice lattice~\cite{RN10877}. This factor is also explained with the thermodynamical approach of Jaccard's theory, but initially the value was two times too high. This problem was solved by Ryzhkin and Whitworth~\cite{RN13115} by using a mean field cluster approximation. Our derivation is just based on the diffusion of defects and so we proved that the $\Phi$ factor can be obtained from kinematic considerations only. No complex thermodynamic calculations are necessary. 

Our approach results in a deeper understanding of the mechanisms behind the second term of the flux density. The polarisation of the ice structure is inducing concentration gradients in the defects that are compensating the ohmic currents. If only one type of charge carrier is present, the electric current induced by a static electric field will completely disappear. Indeed, Eq.~\ref{eq:PlJ} from the macroscopic Maxwell equations provides us with a straightforward way to calculate $\mathbf{P_l}$,

\begin{equation}
\frac{\partial \mathbf{P_l}}{\partial t} = \mathbf{J_b} =  e_{DL}\mathbf{j_{D}}.   \label{eq:PlJN}
\end{equation}
Filling in Eq.~\ref{eq:fluxD4}, this equation can be transformed into a first-order differential equation for $\mathbf{P_l}$

\begin{equation}
\frac{\partial \mathbf{P_l}}{\partial t} +   \frac{\Phi\sigma_D}{e_{DL}^2} \mathbf{P_l} = \sigma_D  \mathbf{E}.  \label{eq:PlJN2}
\end{equation} 
So if a constant electric field is applied, the polarisation density will reach the constant value $\mathbf{P_{lo}}$,

\begin{equation}
\mathbf{P_{lo}} = \frac{e_{DL}^2}{\Phi} \mathbf{E}. \label{eq:PlJN3}
\end{equation} 
Filling Eq.~\ref{eq:PlJN3} into Eq.~\ref{eq:fluxD4} shows that both $\mathbf{J_D}$ and $\mathbf{j_D}$ are zero in steady state. So, the ice becomes polarised when the ohmic current is perfectly compensated by the diffusion current. 

\begin{figure*}
\centering{\includegraphics[width=120mm]{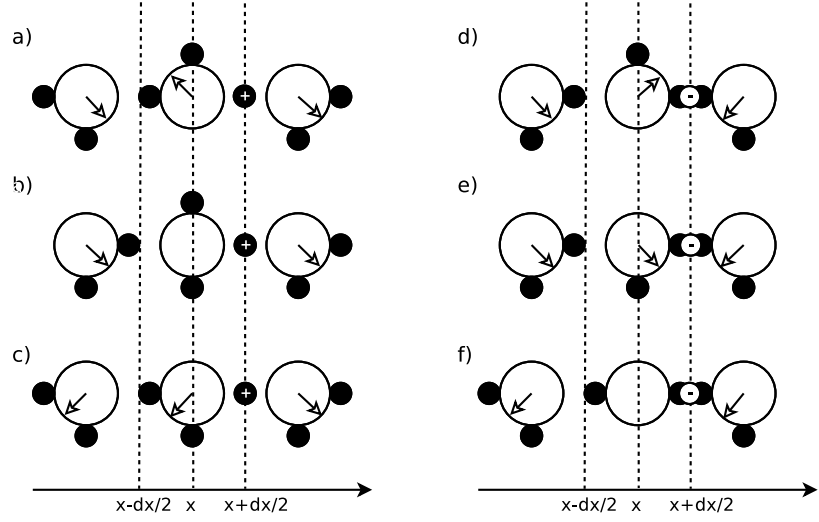}}
\caption{\label{fig:currentions}In a-c H$^+$ ions are drawn when they are located at position $x+dx/2$. Notice that the positive ion of charge $e$ is accompanied with an L defect (charge -$e_{DL}$). In d-f the OH$^-$ ions at position $x+dx/2$ is represented by a D defect (charge $e_{DL}$) and a negative charge of size $-e$.}
\end{figure*}  
 
Experiments show that ice has a non-zero conductivity. This is only possible when different types of defects are opening new paths for each other. The flux equations for all the defects can be derived using a similar approach as for the D defects. For the L defects, this is straightforward. For the ions, a Figure similar to Fig.~\ref{fig:current} must be designed taking into account that the ions are are accompanied by bound charges (see Eq.~\ref{eq:e+5}). In Fig.~\ref{fig:currentions} the two different ions are drawn at position $x+dx/2$. The H$^+$ ion (with charge $e$) is located in an L defect. The OH$^-$ ion is represented in a similar way. A negative charge $-e$ is located in a D defect. Using these representations, equivalent equations for Eq.~\ref{eq:fluxD4} can be derived for all types of defects:

\begin{eqnarray}
\mathbf{j_{+}} &=& \frac{\sigma_{+} }{e_{\pm}^2} (e_{\pm} \mathbf{E} +  \frac{\Phi}{e_{DL}} \mathbf{P_l}), \label{eq:fluxes1} \\ 
\mathbf{j_{-}} &= &\frac{\sigma_{-} }{e_{\pm}^2} (-e_{\pm} \mathbf{E} -  \frac{\Phi}{e_{DL}} \mathbf{P_l}), \label{eq:fluxes2} \\
\mathbf{j_D} &=& \frac{\sigma_D }{e_{DL}^2} (e_{DL} \mathbf{E} -  \frac{\Phi}{e_{DL}} \mathbf{P_l}), \label{eq:fluxes3} \\ 
\mathbf{j_L} &= &\frac{\sigma_L }{e_{DL}^2} (-e_{DL} \mathbf{E} +  \frac{\Phi}{e_{DL}} \mathbf{P_l}). \label{eq:fluxes4} 
\end{eqnarray}  
The diffusion current in the four equations are all dependent on $\mathbf{P_l}$ which is now equal to (see Eq.~\ref{eq:PlJ}), 

\begin{equation}
\frac{\partial \mathbf{P_l}}{\partial t} = \mathbf{J_b} =    e_{DL}(-\mathbf{j_{+}}+\mathbf{j_{-}}+\mathbf{j_D} -\mathbf{j_L}).   \label{eq:PlJ2}
\end{equation} 
The signs of the bound currents are obtained from Table~\ref{Table1} and related to the sign of the bound charges that are transported together with the defects. Eqs.~\ref{eq:fluxes1} to \ref{eq:PlJ2} are describing the electric behaviour in ice and will be examined in the next sections.

\subsection{The susceptibility and Debye relaxation time}

If an oscillating electric field with pulsation $\omega$ is applied, Eq.~\ref{eq:PlJ2} can be rewritten as

\begin{equation}
i \omega \mathbf{P_l} =   e_{DL}(-\mathbf{j_{+}}+\mathbf{j_{-}}+\mathbf{j_D} -\mathbf{j_L}), \label{eq:int2}
\end{equation} 
with $i$ the imaginary unit. If we combine Eq.~\ref{eq:int2} with Eqs.~\ref{eq:fluxes1}-\ref{eq:fluxes4} an analytical solution for $\mathbf{P_l}$ is obtained:

\begin{equation}
\mathbf{P_l} = \frac{e_{DL}\tau_l^i (\frac{\sigma_{D} + \sigma_{L}}{e_{DL}} -\frac{\sigma_{+} +\sigma_{-} }{e_{\pm}}) \mathbf{E}  } {(1+ i\omega \tau_l^i)},  \label{eq:int3}
\end{equation} 
with $\tau_l^i$ the relaxation time of the ice lattice, defined as:

\begin{equation}
\frac{1}{\tau_l^i} = \Phi (\frac{\sigma_++\sigma_{-} }{e_{\pm}^2} +\frac{\sigma_{D}+\sigma_{L}}{e_{DL}^2} ). \label{eq:taul}
\end{equation} 
This relaxation time is the Debye relaxation time, a fundamental property of the lattice quantifying how fast the lattice polarisation will change under influence of an external electric field. The more defects in the lattice the faster the lattice polarisation will be obtained. 

When a static field $\mathbf{E}$ is applied to the ice specimen, the susceptibility of the ice lattice $\chi^{i}_l$ is defined  as  

\begin{equation}
\mathbf{P_{lo}} =  \epsilon_o \chi^{i}_l \mathbf{E}.  \label{eq:PlNew} 
\end{equation}
with $\mathbf{P_{lo}}$ the DC polarisation density. 
So, if we put $\omega= 0$ in Eq.~\ref{eq:int3}, we get

\begin{equation}
\chi_l^i = \frac{e_{DL}(\frac{\sigma_{D} + \sigma_{L}}{e_{DL}} -\frac{\sigma_{+} + \sigma_{-} }{e_{\pm}})  } { \epsilon_o \Phi (\frac{\sigma_++\sigma_{-} }{e_{\pm}^2} +\frac{\sigma_{D}+\sigma_{L}}{e_{DL}^2} )}, \label{eq:chili}
\end{equation} 
In the limit that $\sigma_{D}+\sigma_{L}  \gg \sigma_{+}+\sigma_{-}$, the electric susceptibility 

\begin{equation}
\chi_l^i = \frac{e_{DL}^2}{\epsilon_o \Phi}. \label{eq:chili2}
\end{equation} 
However, if the opposite is true $ \sigma_{+}+\sigma_{-} \gg \sigma_{D}+\sigma_{L} $ we obtain a negative susceptibility

\begin{equation}
\chi_l^i = - \frac{e_{DL}e_{\pm}}{\epsilon_o \Phi}. \label{eq:chili3}
\end{equation} 
Although strange at first sight, this is an obvious result taking into consideration that now the movement of ions is polarising the lattice. As can be seen in Fig.~\ref{fig:2Dmodel2}a, the movement of ions results in an lattice of water molecules oppositely polarised to the applied electric field.   

It is also interesting to bring $\chi_l^i$ and $\tau_l^i$ in an alternative formulation of Eq.~\ref{eq:PlJ2},  

\begin{equation}
\frac{\partial \mathbf{P_l}}{\partial t} = \frac{\mathbf{P_{lo}} - \mathbf{P_l}}{\tau_l^i}  = \frac{\epsilon_o \chi_l^i \mathbf{E} -\mathbf{P_l}}{\tau_l^i}, \label{eq:PlJ3}
\end{equation}  
which is a first-order differential equation for $\mathbf{P_l}$ with a time constant $\tau_l^i$ and a DC polarisation density $\mathbf{P_{lo}}$. 

\subsection{The DC and high frequency conductivity}

The electrical current density is the sum of the current density of the bound and the free charges~\cite{RN13018}. Applying the results from Table~\ref{Table1},  

\begin{equation}
\mathbf{ J} =  \mathbf{ J_b} +  \mathbf{ J_f} = e_{DL}(-\mathbf{j_{+}}+\mathbf{j_{-}}+\mathbf{j_D} -\mathbf{j_L}) + e(\mathbf{j_{+}}-\mathbf{j_{-}}).  \label{eq:J}
\end{equation}
Using Eqs.~\ref{eq:fluxes1}, \ref{eq:fluxes2}, \ref{eq:PlJ2} and \ref{eq:PlJ3}, this equation can also be rewritten as

\begin{equation}
\mathbf{ J} =   \frac{\mathbf{P_{lo}} - \mathbf{P_l}}{\tau_l^i}  +  e(\frac{\sigma_{+} + \sigma_{-}}{e_{\pm}} \mathbf{E} +  \frac{\Phi}{e_{DL}} \frac{\sigma_{+}+ \sigma_{-} }{e_{\pm}^2}  \mathbf{P_l}).
 \label{eq:J2}
\end{equation}
We are interested in this equation in two limiting conditions. The first one is the high frequency condition, where $\mathbf{P_l}$ reaches 0 (Eq.~\ref{eq:int3}). Using Eqs.~\ref{eq:taul}, \ref{eq:PlNew} and \ref{eq:chili}, the electrical current density simplifies to 

\begin{equation}
\mathbf{ J} =  \sigma_{\infty} \mathbf{E} = (\sigma_{+} + \sigma_{-} +\sigma_{D}+\sigma_{L} ) \mathbf{E},
 \label{eq:J3}
\end{equation} 
with $\sigma_{\infty}$ the high frequency conductivity. This high frequency limit of the conductivity can be interpreted as a parallel network of the defects~\cite{RN10943}. In this parallel network is the total conductivity the sum of the conductivities of every defect. The resulting value of $\sigma_{\infty}$ will be mainly determined by the highest values of the conductivity. 

The other limiting condition defines the DC conductivity $\sigma_{o}$. If $\omega$ = 0, $\mathbf{P_{l}} = \mathbf{P_{lo}}$, and Eq.~\ref{eq:J2} reduces to

\begin{equation}
\mathbf{ J} =  \sigma_{o} \mathbf{E} =  e(\frac{\sigma_{+} + \sigma_{-}}{e_{\pm}} \mathbf{E} +  \frac{\Phi}{e_{DL}} \frac{\sigma_{+}+ \sigma_{-} }{e_{\pm}^2}  \mathbf{P_{lo}}) \label{eq:J3}
\end{equation}
With some straightforward calculations using Eq.~\ref{eq:PlNew} and Eq.~\ref{eq:chili}, one obtains the following equation for $\sigma_{o}$ 

\begin{equation}
\frac{e^2}{\sigma_{o}} = \frac{e_\pm^2}{\sigma_{+}+\sigma_{-}} + \frac{e_{DL}^2}{\sigma_{D}+\sigma_{L}}. \label{eq:sDC}
\end{equation}  
The DC conductivity is interpreted as a serial netwerk of both the Bjerrum and ionic defects weighted by the square of their charge~\cite{RN10943}. The DC conductivity will be determined by the smallest values of conductivity in the network. This result can be understood using Fig.~\ref{fig:2Dmodel2}. Both the Bjerrum and ionic defects are closing their own paths hopping through the ice structure, so they need each other to keep their paths open. Even when one type of defect is abundantly present, they cannot contribute to the DC conductivity without the help of the other type of defects.

\section{Discussion}

\subsection{The use of the configuration vector}
We proved that no configuration vector is necessary to build a straightforward version of Jaccard's theory. There is no problem in using the configuration vector, but is important that it is done correctly taking into account the difference between bound and free charges. 
 
Petrenko and Withworth define the configuration vector as~\cite{RN10943}: 'An important parameter in describing the state of the ice is the number of chains of bonds crossing a given plane that are oriented in one direction or the other.' This quantity is described by the vector $\mathbf{ \Omega}$ equal to

\begin{equation}
\mathbf{ \Omega}(t) = \int_{0}^{t} [ \mathbf{j_{+}}(t') - \mathbf{j_{-}}(t') - \mathbf{j_{D}}(t') - \mathbf{j_{L}}(t') ] dt'. \label{eq:omega1}
\end{equation}
Comparison of this equation with Eq.~\ref{eq:PlJ2} makes it clear that the configuration vector is just another formulation of $\mathbf{P_l}$, i.e.

\begin{equation}
\mathbf{P_l} = - e_{DL}\mathbf{ \Omega}(t), \label{eq:omega2}
\end{equation}
so it is completely unnecessary to define. Petrenko and Withworth\cite{RN10943} use this relation between $\mathbf{ \Omega}$  and $\mathbf{P_l}$ to calculate the susceptibility in the case of one kind of defect (section 4.5.1), but for the more general formulations (section 4.5.2) $\mathbf{P_l}$ is calculated from 

\begin{equation}
\frac{\partial \mathbf{P_l}}{\partial t} = \mathbf{J},   \label{eq:Plwrong}
\end{equation} 
 where $\mathbf{J}$ is the total current density instead of the current density of bound charges, as was defined in the Maxwell equations. This leads to a wrong equation for the susceptibility, given by 

\begin{equation}
\chi_l^i = \frac{((\sigma_{+}+\sigma_{-})/e_\pm - (\sigma_{D}+\sigma_{L})/e_{DL})^2}{\epsilon_o \Phi ((\sigma_{+}+\sigma_{-})/e_\pm^2 + (\sigma_{D}+\sigma_{L})/e_{DL}^2)^2}, \label{eq:chiwrong}
\end{equation} 
For $\sigma_{D}+\sigma_{L}  \gg \sigma_{+}+\sigma_{-}$, this equation also reduces to Eq.~\ref{eq:chili2}, but for $ \sigma_{+}+\sigma_{-} \gg \sigma_{D}+\sigma_{L} $ the susceptibility is positive

\begin{equation}
\chi_l^i = \frac{e_{\pm}^2}{\epsilon_o \Phi}, \label{eq:chiwrong2}
\end{equation} 
opposite to what we expect. Indeed, the ions have a bound charge with a sign opposite to their free charge (see Table~\ref{Table1}) and should therefore polarise the lattice opposite to the electric field. 
  
The configuration vector can become a parameter that only has to be introduced to understand the older literature about ice. When introduced, one should emphasize the link between this vector and the polarisation density (Eq.~\ref{eq:omega2}) and the fact that in the older literature an incorrect formula for the polarisation density was used. 

\subsection{The impact of the new approach}
It was our own confusion with several aspects of Jaccard's theory that was the motivation for this new formulation of Jaccard's theory. Therefore, we are convinced that this paper will have a didactical effect. That's also the reason why we have chosen for a detailed formulation of the new approach instead of focussing on the isolated differences. A person not familiar with the physics of ice should be able to follow the whole reasoning. The differentiation of bound and free charges will help students and researchers to understand the physics of ice more easily and to link it to their standard knowledge of electromagnetism.  

Our kinetic approach for deriving the flux densities of the defects leads to the correct experimental value of $\Phi$. This factor can also be derived using an thermodynamic approach calculating the entropy production produced by the movement of defects, but in a much more complex way~\cite{RN10877,RN13115}. Petrenko mentions that Jaccard got stuck in his kinetic approach and therefore turned to the thermodynamic approach~\cite{RN10943}. A crucial element in the development of our kinetic theory was the idea that the $\Phi$ terms of Jaccard's theory were classical diffusion terms of the defects. The fact that the kinetic approach of Jaccard was not leading to the correct equations was puzzling and suggesting that some things were not completely understood.    

For pure ice, the impact of the new equation for the susceptibility (Eq.~\ref{eq:chili} instead of Eq.~\ref{eq:chiwrong}) is low. Both equations give the same result if $\sigma_{D}+\sigma_{L}  \gg \sigma_{+}+\sigma_{-}$. For pure ice $\chi_l^i$ becomes systematically larger over a wide temperature range (from 273 until 140 K)~\cite{RN13110}, which can only be correct if the DL defects dominate over the whole temperature range. This is confirmed experimentally by Camplin~\cite{RN11023}. 

In doped ice, the concentration of $ \sigma_{+}+\sigma_{-}$ can be increased artificially~\cite{RN10943}. Camplin at al examined HF doped ice~\cite{RN11023}. They estimated the conductivity of the different defects and find that at lower temperatures the conductivity of the H$^+$ ions is significantly higher than for the DL defects. If their assumptions are correct, this should lead to a negative susceptibility. However, no experimental results of the susceptibility are mentioned. Takei and Maeno investigated HCL doped ice~\cite{RN13137} and reported significant decreases in the value of the susceptibility at low temperatures. However, no negative values are found. In this low temperature range, they found that $\sigma_\infty$ decreases to a value a little higher than $\sigma_o$. Using Eqs.~\ref{eq:J2} and \ref{eq:sDC}, it  can be easily derived that $\sigma_\infty \approx \sigma_o$ is only possible when 

\begin{equation}
\frac{\sigma_{DL}}{e_{DL}} \approx \frac{\sigma_\pm}{e_\pm}. \label{eq:final}
\end{equation}
       This result proves that the electric susceptibility (Eq.~\ref{eq:chili}) should decrease without becoming negative. Taking Eq.~\ref{eq:final} into account and the fact that in all the data of doped ice we found  $\sigma_\infty > \sigma_o$ (also in the data of Camplin et al.), we doubt that a negative susceptibility is possible to obtain using doping.

\section{Conclusions}
In most versions of Jaccard's theory for the electric properties in ice the difference between bound charges and free charges is ignored. Using the macroscopic Maxwell equations this imperfection is overcome in a new formulation of Jaccard's theory using the polarisation density instead of the the ad hoc 'configuration vector'. We used a kinematic approach resulting in a $\Phi$ factor equal to the experimental factor, something Jaccard did not find. We also prove that within the old formulation a wrong formula for the electric susceptibility is obtained. An improved formula is proposed offering a challenge to reconsider the susceptibility of doped ice at low temperatures.

\section*{Acknowledgements} 
The author thanks Ward De Jonghe, Prof. Jacques Tempere and Prof. Em. Roland Van Meirhaeghe for their support and their critical reviews of this work.



\bibliography{literatuur}

\nolinenumbers

\end{document}